\documentclass[aps,prd,preprint,superscriptaddress,tightenlines,%
nofootinbib]{revtex4}
\newcommand{\PRE}[1]{{#1}}   % Use if preprint style

\usepackage{amsmath,amssymb}
\usepackage{bm}
\usepackage{epsfig}
\usepackage{latexsym}

\newcommand{\postscript}[2]{\setlength{\epsfxsize}{#2\hsize}
   \centerline{\epsfbox{#1}}}
\newcommand{\comment}[1]{}
\def\thalf{\tfrac{1}{2}}

\def\comment#1{{}}

\def\tfrac#1#2{{\textstyle\frac{#1}{#2}}}
\def\slashchar#1{\setbox0=\hbox{$#1$}           % set a box for #1
   \dimen0=\wd0                                 % and get its size
   \setbox1=\hbox{/} \dimen1=\wd1               % get size of /
   \ifdim\dimen0>\dimen1                        % #1 is bigger
      \rlap{\hbox to \dimen0{\hfil/\hfil}}      % so center / in box
      #1                                        % and print #1
   \else                                        % / is bigger
      \rlap{\hbox to \dimen1{\hfil$#1$\hfil}}   % so center #1
      /                                         % and print /
   \fi}
\newif\ifnref

\nreffalse

\input epsf
\def\figin{\epsfcheck\figin}\def\figins{\epsfcheck\figins}
\def\epsfcheck{\ifx\epsfbox\UnDeFiNeD
\message{(NO epsf.tex, FIGURES WILL BE IGNORED)}
\gdef\figin##1{\vskip2in}\gdef\figins##1{\hskip.5in}% blank space instead
\else\message{(FIGURES WILL BE INCLUDED)}%
\gdef\figin##1{##1}\gdef\figins##1{##1}\fi}
\def\DefWarn#1{}
\def\figinsert{\goodbreak\midinsert}
\def\ifig#1#2#3{\DefWarn#1\xdef#1{fig.~\the\figno}
\writedef{#1\leftbracket fig.\noexpand~\the\figno}%
\figinsert\figin{\centerline{#3}}\medskip\centerline{\vbox{\baselineskip12pt
\advance\hsize by -1truein\noindent\footnotefont{\bf Fig.~\the\figno } #2}}
\bigskip\endinsert\global\advance\figno by1}

\def\hat{\widehat}
\begin{document}

\preprint{
\hfil
\begin{minipage}[t]{3in}
\begin{flushright}
\vspace*{.4in}
%hep-ph/yymmnnn
\end{flushright}
\end{minipage}
}

\title{Direct photons as probes of low mass strings at the LHC
%\PRE{\vspace*{1.5in}}
\PRE{\vspace*{0.3in}} }

\author{Luis A. Anchordoqui}
\affiliation{Department of Physics,\\
University of Wisconsin-Milwaukee,
 Milwaukee, WI 53201, USA
\PRE{\vspace*{.1in}}
}

\author{Haim Goldberg}
\affiliation{Department of Physics,\\
Northeastern University, Boston, MA 02115, USA
\PRE{\vspace*{.1in}}
}

\author{Satoshi Nawata}
\affiliation{Department of Physics,\\
University of Wisconsin-Milwaukee,
 Milwaukee, WI 53201, USA
\PRE{\vspace*{.1in}}
}

\author{Tomasz R. Taylor}
\affiliation{Department of Physics,\\
Northeastern University, Boston, MA 02115, USA
\PRE{\vspace*{.1in}}
}

\date{April 2008}
\PRE{\vspace*{.5in}}
\begin{abstract}
\vskip 3mm
\noindent The LHC program will include the identification of events
with single prompt high-$k_\perp$ photons as probes of new physics. We
show that this channel is uniquely suited to search for experimental
evidence of TeV-scale open string theory.  At the parton level, we
analyze single photon production in gluon fusion, $gg \to \gamma g$,
with open string states propagating in intermediate channels. If the
photon mixes with the gauge boson of the baryon number, which is a
common feature of D-brane quivers, the amplitude appears already at
the string disk level. It is completely determined by the mixing
parameter (which is actually determined in the minimal theory) -- and
it is otherwise model-(compactification-) independent. We discuss
the string signal cross sections as well
as the QCD background.  The present analysis takes into account the
recently obtained decay widths of first Regge recurrences, which are
necessary for more precise determination of these cross sections in
the resonant region. A vital part of the background discussion
concerns the minimization of misidentified $\pi^0$'s emerging from
high-$p_\perp$ jets. We show that even for relatively small mixing,
100~fb$^{-1}$ of LHC data could probe deviations from standard model
physics associated with TeV-scale strings at a $5\sigma$ significance,
for $M_{\rm string}$ as large as 2.3~TeV.  It is also likely that
resonant bumps could be observed with approximately the same
signal-to-noise ratio.
\end{abstract}

\maketitle

\section{General Idea}

The CERN's Large Hadron Collider (LHC) is the greatest basic science
endeavor in history. Spectacular physics results are expected to
follow in short order once it turns on this year. LHC will push
nucleon-nucleon center-of-mass energies up to $\sqrt{s} = 14~{\rm
  TeV}$ for $pp$ collisions and $\sqrt{s} = 5.5~{\rm TeV}$ for Pb-Pb
collisions.  The ATLAS and CMS detectors will observe the
highest-energy particle collisions produced by the accelerator,
whereas the ALICE detector will observe the very messy debris of heavy
ion collisions.  The LHC will probe deeply into the sub-fermi
distances, committing to careful searches for new particles and
interactions at the TeV scale.

At the time of its formulation and for years thereafter, Superstring
Theory was regarded as a unifying framework for Planck-scale quantum
gravity and TeV-scale Standard Model (SM) physics. Important advances
were fueled by the realization of the vital role played by D-branes
\cite{joe} in connecting string theory to phenomenology
\cite{reviews}. This has permitted the formulation of string theories
with compositeness setting in at TeV scales~\cite{Antoniadis:1998ig}
and large extra dimensions. There are two paramount phenomenological
consequences for TeV scale D-brane string physics: the emergence of
Regge recurrences at parton collision energies $\sqrt{\hat s} \sim
{\rm string\ scale} \equiv M_s;$ and the presence of one or more
additional $U(1)$ gauge symmetries, beyond the $U(1)_Y$ of the SM.
The latter follows from the property that the gauge group for open
strings terminating on a stack of $N$ identical D-branes is $U(N)$
rather than $SU(N)$ for $N>2.$ (For $N=2$ the gauge group can be
$Sp(1)$ rather than $U(2)$.) In this paper we exploit both these
properties in order to obtain a ``new physics'' signal at LHC which,
if traced to low scale string theory, could with 100 fb$^{-1}$ of data
probe deviations from SM physics at a $5\sigma$ significance for $M_s$
as large as 2.3~TeV. A short version highlighting the salient results
of our analysis has been issued as a companion
Letter~\cite{Anchordoqui:2007da}. The present analysis, however, takes
into account the recently obtained decay widths of first Regge
recurrences~\cite{widths}, which are necessary for more precise
determination of cross sections in the resonant region.

To develop our program in the simplest way, we will work within the
construct of a minimal model in which we consider scattering processes
which take place on the (color) $U(3)$ stack of D-branes. In the
bosonic sector, the open strings terminating on this stack contain, in
addition to the $SU(3)$ octet of gluons, an extra $U(1)$ boson
($C_\mu$, in the notation of~\cite{Berenstein:2006pk}), most simply
the manifestation of a gauged baryon number symmetry. The $U(1)_Y$
boson $Y_\mu$, which gauges the usual electroweak hypercharge
symmetry, is a linear combination of $C_\mu$, the $U(1)$ boson $B_\mu$
terminating on a separate $U(1)$ brane, and perhaps a third additional
$U(1)$ (say $W_\mu$) sharing a $U(2)$ brane to which are also a
terminus for the $SU(2)_L$ electroweak gauge bosons $W_\mu^a.$ Thus,
critically for our purposes, the photon $A_\mu$, which is a linear
combination of $Y_\mu$ and $W^3_\mu$ {\em will participate with the
  gluon octet in (string) tree level scattering processes on the color
  brane, processes which in the SM occur only at one-loop level.} Such
a mixing between hypercharge and baryon number is a generic property
of D-brane quivers, see {\it e.g}.\
Refs.\cite{ant,bo,Berenstein:2006pk}. The vector boson $Z'_\mu$, 
orthogonal to the hypercharge, must grow a mass $M_{Z'}$ in order to 
avoid long range forces between baryons other than gravity and Coulomb 
forces. 
%The anomalous mass growth allows the survival of global 
%baryon number conservation, 
%preventing fast proton decay~\cite{Ghilencea:2002da}.

The process we consider (at the parton level) is $gg\rightarrow
g\gamma$, where $g$ is an $SU(3)$ gluon and $\gamma$ is the photon. As
explicitly calculated below, this will occur at string disk (tree)
level, and will be manifest at LHC as a non-SM contribution to
$pp\rightarrow \gamma +\ {\rm jet}$.  A very important property of
string disk amplitudes is that they are completely model-independent;
thus the results presented below are robust, because {\em they hold
  for arbitrary compactifications of superstring theory from ten to
  four dimensions, including those that break supersymmetry}.  The SM
background for this signal originates in the parton tree level
processes $g q \rightarrow \gamma q,\ g\bar q\rightarrow \gamma\bar
q,\ {\rm and } \ q\bar q\rightarrow \gamma g$. Of course, the SM
processes will also receive stringy corrections which should be added
to the pure bosonic contribution as part of the
signal~\cite{Cullen:2000ef,Burikham:2004su,Meade:2007sz,Domokos:1998ry}.
We leave this evaluation to a subsequent publication~\cite{ws}; thus,
the contribution from the bosonic process calculated here is to be
regarded as a lower bound to the stringy signal. It should also be
stated that, in what follows, we do not include effects of
Kaluza-Klein recurrences due to compactification. We assume that all
such effects are in the gravitational sector, and hence occur at
higher order in string coupling~\cite{Cullen:2000ef}.

\begin{figure}[tbp]
\postscript{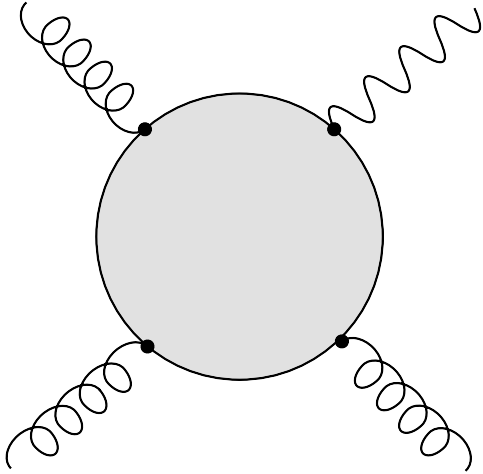}{0.4}
\caption{Open string disk diagram for $gg \to g \gamma$ scattering.  The
  dots represent vertex insertions of the gauge bosons on the boundary
  of the world sheet.}
\label{diag}
\end{figure}

The plan of the paper is as follows. In the next section
we outline the calculation of the string amplitude for the process
pictured in Fig.~\ref{diag} and show that there is no amplitude
containing the zero mass poles of the SM. In Sec.~\ref{s3} we first
calculate cross sections for gluon fusion in the resonance region as
well as QCD background, for a simple $k_{\perp, {\rm min}}$ cut on the
transverse momentum of the photon. A vital part of the background
discussion concerns the minimization of misidentified $\pi^0$'s from
high-$p_\perp$ jets.  After that we calculate the signal-to-noise
ratio and show that a significant deviation from SM can be obtained
for $k_{\perp, {\rm min}} > 300$~GeV for $M_s$ as large a 2.3~TeV.  In
Sec.~\ref{s4} we delineate the search for resonant structure in the
data. Our conclusions are collected in Sec.~\ref{s5}. The Appendices contain 
some additional formulae referred to in the main text.

\section{The String Amplitude}
\label{s2}

The most direct way to compute the amplitude for the scattering of
four gauge bosons is to consider the case of polarized particles
because all non-vanishing contributions can be then generated from a
single, maximally helicity violating (MHV), amplitude -- the so-called
{\it partial\/} MHV amplitude~\cite{ptmhv}.  Assume that two vector bosons,
with the momenta $k_1$ and $k_2$, in the $U(N)$ gauge group states
corresponding to the generators $T^{a_1}$ and $T^{a_2}$ (here in the
fundamental representation), carry negative helicities while the other
two, with the momenta $k_3$ and $k_4$ and gauge group states $T^{a_3}$
and $T^{a_4}$, respectively, carry positive helicities. (All
momenta are incoming.)  Then the
partial amplitude for such an MHV configuration is given by~\cite{STi,STii}
\begin{equation}
\label{ampl}
A(1^-,2^-,3^+,4^+) ~=~ 4\, g^2\, {\rm Tr}
  \, (\, T^{a_1}T^{a_2}T^{a_3}T^{a_4}) {\langle 12\rangle^4\over
    \langle 12\rangle\langle 23\rangle\langle 34\rangle\langle
    41\rangle}V(k_1,k_2,k_3,k_4)\ ,
\end{equation}
where $g$ is the $U(N)$ coupling constant, $\langle ij\rangle$ are the
standard spinor products written in the notation of
Refs.~\cite{Mangano,Dixon}, and the Veneziano formfactor,
\begin{equation}
\label{formf}
V(k_1,k_2,k_3,k_4)=V(s,t,u)= {\Gamma(1-s)\ \Gamma(1-u)\over
    \Gamma(1+t)}\ ,
\end{equation}
is the function of Mandelstam variables, here
normalized in the string units:
\begin{equation}
\label{mandel}
s={2k_1k_2\over M_s^2},~ t={2
  k_1k_3\over M_s^2}, ~u={2 k_1k_4 \over M_s^2}:\qquad s+t+u=0.
\end{equation}
(For simplicity we drop carets for the parton subprocess.)
Its low-energy
  expansion reads
\begin{equation}
\label{vexp}
V(s,t,u)\approx 1-{\pi^2\over 6}s\,
    u-\zeta(3)\,s\, t\, u+\dots
\end{equation}

We first consider the amplitude involving three $SU(N)$ gluons
$g_1,~g_2,~g_3$ and one $U(1)$ gauge boson $\gamma_4$ associated to
the same $U(N)$ quiver:
\begin{equation}
\label{gens}
T^{a_1}=T^a \ ,~ \ T^{a_2}=T^b\ ,~ \
  T^{a_3}=T^c \ ,~ \ T^{a_4}=QI\ ,
\end{equation}
where $I$ is the $N{\times}N$ identity matrix and $Q$ is the
$U(1)$ charge of the fundamental representation. The $U(N)$
generators are normalized according to
\begin{equation}
\label{norm}
{\rm Tr}(T^{a}T^{b})={1\over 2}\delta^{ab}.
\end{equation}
Then the color
factor \begin{equation}\label{colf}{\rm
    Tr}(T^{a_1}T^{a_2}T^{a_3}T^{a_4})=Q(d^{abc}+{i\over 4}f^{abc})\ ,
\end{equation}
where the totally symmetric symbol $d^{abc}$ is the symmetrized trace
while $f^{abc}$ is the totally antisymmetric structure constant.

The full MHV amplitude can be obtained~\cite{STi,STii} by summing
the partial amplitudes (\ref{ampl}) with the indices permuted in the
following way: \begin{equation}
\label{afull} {\cal M}(g^-_1,g^-_2,g^+_3,\gamma^+_4)
  =4\,g^{2}\langle 12\rangle^4 \sum_{\sigma } { {\rm Tr} \, (\,
    T^{a_{1_{\sigma}}}T^{a_{2_{\sigma}}}T^{a_{3_{\sigma}}}T^{a_{4}})\
    V(k_{1_{\sigma}},k_{2_{\sigma}},k_{3_{\sigma}},k_{4})\over\langle
    1_{\sigma}2_{\sigma} \rangle\langle
    2_{\sigma}3_{\sigma}\rangle\langle 3_{\sigma}4\rangle \langle
    41_{\sigma}\rangle }\ ,
\end{equation}
where the sum runs over all 6 permutations $\sigma$ of $\{1,2,3\}$ and
$i_{\sigma}\equiv\sigma(i)$. Note that in the effective field theory
of gauge bosons there are no Yang-Mills interactions that could
generate this scattering process at the tree level. Indeed, $V=1$ at
the leading order of Eq.(\ref{vexp}) and the amplitude vanishes
due to the following identity:
\begin{equation}\label{ymlimit}
{1\over\langle 12\rangle\langle
      23\rangle\langle 34\rangle\langle
      41\rangle}+{1\over\langle 23\rangle\langle
      31\rangle\langle 14\rangle\langle 42\rangle}+{1\over\langle 31\rangle\langle
      12\rangle\langle 24\rangle\langle 43\rangle} ~=~0\ .
\end{equation}
Similarly,
the antisymmetric part of
the color factor (\ref{colf}) cancels out in the full amplitude (\ref{afull}). As a result,
one obtains:
\begin{equation}\label{mhva}
{\cal
    M}(g^-_1,g^-_2,g^+_3,\gamma^+_4)=8\, Q\, d^{abc}g^{2}\langle
  12\rangle^4\left({\mu(s,t,u)\over\langle 12\rangle\langle
      23\rangle\langle 34\rangle\langle
      41\rangle}+{\mu(s,u,t)\over\langle 12\rangle\langle
      24\rangle\langle 13\rangle\langle 34\rangle}\right),
\end{equation}
 where
\begin{equation}
\label{mudef}
\mu(s,t,u)= \Gamma(1-u)\left( {\Gamma(1-s)\over
      \Gamma(1+t)}-{\Gamma(1-t)\over \Gamma(1+s)}\right) .
\end{equation}
All
non-vanishing amplitudes can be obtained in a similar way. In
particular,
\begin{equation}
\label{mhvb}
{\cal M}(g^-_1,g^+_2,g^-_3,\gamma^+_4)=8\, Q\,
  d^{abc}g^{2}\langle 13\rangle^4\left({\mu(t,s,u)\over\langle
      13\rangle\langle 24\rangle\langle 14\rangle\langle
      23\rangle}+{\mu(t,u,s)\over\langle 13\rangle\langle
      24\rangle\langle 12\rangle\langle 34\rangle}\right),
\end{equation}
and the
remaining ones can be obtained either by appropriate permutations or
by complex conjugation.

In order to obtain the cross section for the (unpolarized) partonic
subprocess $gg\to g\gamma$, we take the squared moduli of individual
amplitudes, sum over final polarizations and colors, and average over
initial polarizations and colors. As an example, the modulus square of
the amplitude (\ref{afull}) is:
\begin{equation}
\label{mhvsq}
|{\cal
    M}(g^-_1,g^-_2,g^+_3,\gamma^+_4)|^2=64\, Q^2\, d^{abc}d^{abc}g^{4}
  \left|{s\mu(s,t,u)\over u}+{s\mu(s,u,t)\over t} \right|^2 \, .
\end{equation}
 Taking
into account all $4(N^2-1)^2$ possible initial polarization/color
configurations and the formula~\cite{groupf}
\begin{equation}
\label{dsq}
\sum_{a,b,c}d^{abc}d^{abc}={(N^2-1)(N^2-4)\over 16 N},
\end{equation}
 we
obtain the average squared amplitude
\begin{equation}
\label{mhvav}
|{\cal M}(gg\to
  g\gamma)|^2=g^4Q^2C(N)\left\{ \left|{s\mu(s,t,u)\over
        u}+{s\mu(s,u,t)\over t} \right|^2+(s\leftrightarrow
    t)+(s\leftrightarrow u)\right\},
\end{equation}
 where
\begin{equation}\label{cnn}
C(N)={2(N^2-4)\over N(N^2-1)}.
\end{equation}

The two most interesting energy regimes of $gg\to g\gamma$
scattering are far below the string mass scale $M_s$
and near the threshold for the production of massive string
excitations. At low energies, Eq.~(\ref{mhvav}) becomes
\begin{equation}
\label{mhvlow}
|{\cal M}(gg\to
  g\gamma)|^2\approx g^4Q^2C(N){\pi^4\over 4}(s^4+t^4+u^4)\qquad
  (s,t,u\ll 1) \, .
\end{equation}
The absence of massless poles, at $s=0$ {\it etc.\/}, translated
into the terms of effective field theory, confirms that there are
no exchanges of massless particles contributing to this process.
On the other hand, near the string threshold $s\approx M_s^2$
(where we now restore the string scale)
\begin{equation}
\label{mhvlow3}
|{\cal M}(gg\to g\gamma)|^2\approx
4g^4Q^2C(N){M_s^8+t^4+u^4\over M_s^4(s-M_s^2)^2}
\qquad (s\approx M_s^2).
\end{equation}
The singularity at $s=M^2_s$ needs softening to a Breit-Wigner form, reflecting the finite decay widths of resonances propagating in the $s$ channel. Due to averaging over initial polarizations, Eq.(\ref{mhvlow3}) contains additively contributions from both spin $J=0$ and spin $J=2$ gluonic Regge recurrences
($G^*$ in the notation of Ref.\cite{widths}), created by the incident gluons in the helicity configurations ($\pm \pm$) and ($\pm \mp$), respectively.
The $M_s^8$ term
in Eq.~(\ref{mhvlow3}) originates from $J=0$, and the $t^4+ u^4$ piece
reflects $J=2$ activity. Since the resonance widths are spin-dependent \cite{widths}:
\begin{eqnarray}
\Gamma^{J=0} &=& \frac{3}{4}\alpha_sM_s~ \approx 75 \, (M_s/{\rm TeV})~{\rm GeV}~,\nonumber\\
\Gamma^{J=2} &=& \frac{9}{20}\alpha_sM_s\approx 45 \, (M_s/{\rm TeV})~{\rm GeV}~,
\end{eqnarray}
the pole term (\ref{mhvlow3}) should be smeared as
\begin{equation}
\label{mhvlow2}
|{\cal M}(gg\to g\gamma)|^2\simeq
\frac{4g^4Q^2C(N)}{M_s^4}\bigg[{M_s^8\over (s-M_s^2)^2+(\Gamma^{J=0} M_s)^2}+
{t^4+u^4\over (s-M_s^2)^2+(\Gamma^{J=2} M_s)^2}\bigg].
\end{equation}

In what follows we will take $N=3$ and set $g$ equal to the QCD coupling
constant, $\alpha_s = (g^2/4\pi) \sim 0.1$. Before
proceeding with numerical calculation, we need to make precise the
value of $Q$. If we were considering the process $gg\rightarrow C^0 g,$
where $C^0$ is the $U(1)$ gauge field tied to the $U(3)$ brane, then $Q =
\sqrt{1/6}$ due to the normalization condition~(\ref{norm}). However,
for $gg\rightarrow \gamma g$ there are two additional projections:
from $C_\mu$ to the hypercharge boson $Y_\mu$, giving a mixing factor
$\kappa$; and from $Y_\mu$ onto a photon, providing an additional
factor $\cos\theta_W \ (\theta_W=$ Weinberg angle). The $C^0-Y$ mixing
coefficient is model dependent: in the minimal
model~\cite{Berenstein:2006pk} it is quite small, around $\kappa
\simeq 0.12$ for couplings evaluated at the $Z$ mass, which is
modestly enhanced to $\kappa \simeq 0.14$ as a result of RG running of
the couplings up to 2.5~TeV.  It should be noted that in
models~\cite{ant,bo} possessing an additional $U(1)$ which partners
$SU(2)_L$ on a $U(2)$ brane, the various assignment of the charges can
result in values of $\kappa$ which can differ considerably from
$0.12.$ In what follows, we take as a fiducial value $\kappa^2 =
0.02.$ Thus, if (\ref{mhvlow2}) is to describe $gg\rightarrow \gamma
g,$ we modify our definition of $Q$ given in Eq.~(\ref{gens}) to
accommodate the additional mixings, and obtain
\begin{equation}
Q^2= \tfrac{1}{6} \ \kappa^2 \ \cos^2\theta_W \simeq 2.55\times
10^{-3}\ \left(\kappa^2/0.02\right)\ \ .
\label{Q2}
\end{equation}
In the remainder of the paper, we explore potential searches for
 Regge excitations of fundamental strings at LHC.

\section{Isolated hard photons}
\label{s3}

\begin{figure}[tbp]
\postscript{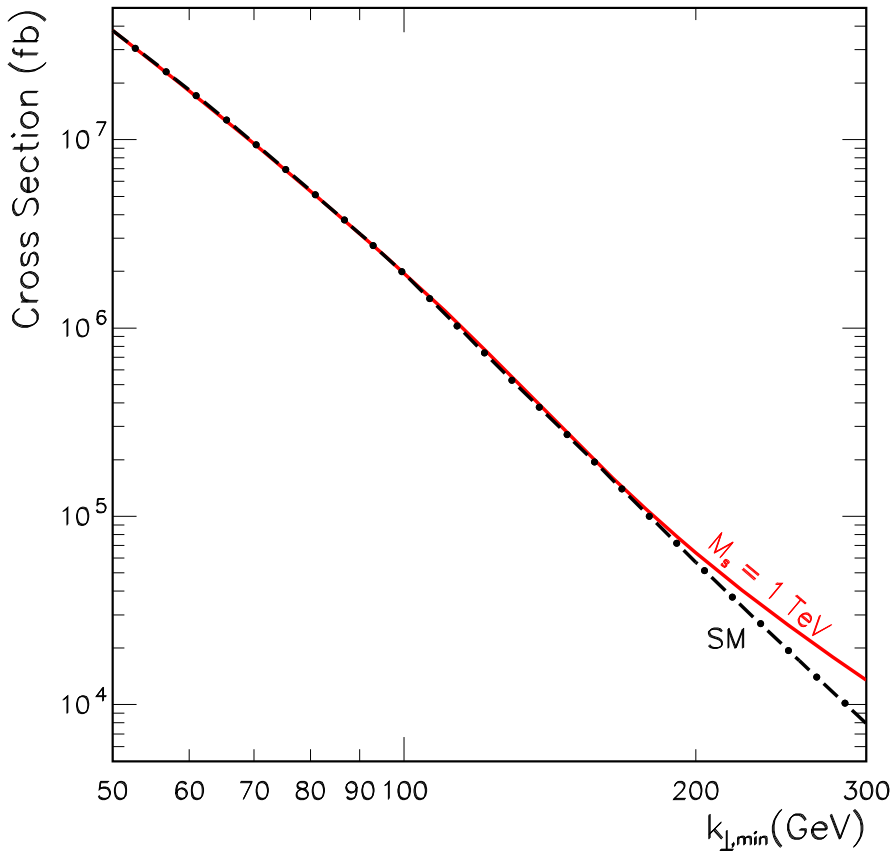}{0.8}
\caption{Behavior of the QCD cross section for $pp \to \gamma + {\rm
    jet}$ (dot-dashed line) as a function of $k_{\perp, {\rm min}}$.
  The string cross section overlying the QCD background is also shown
  as a solid line, for $M_s = 1$~TeV.}
\label{fig:ptfig}
\end{figure}

\begin{figure}
\postscript{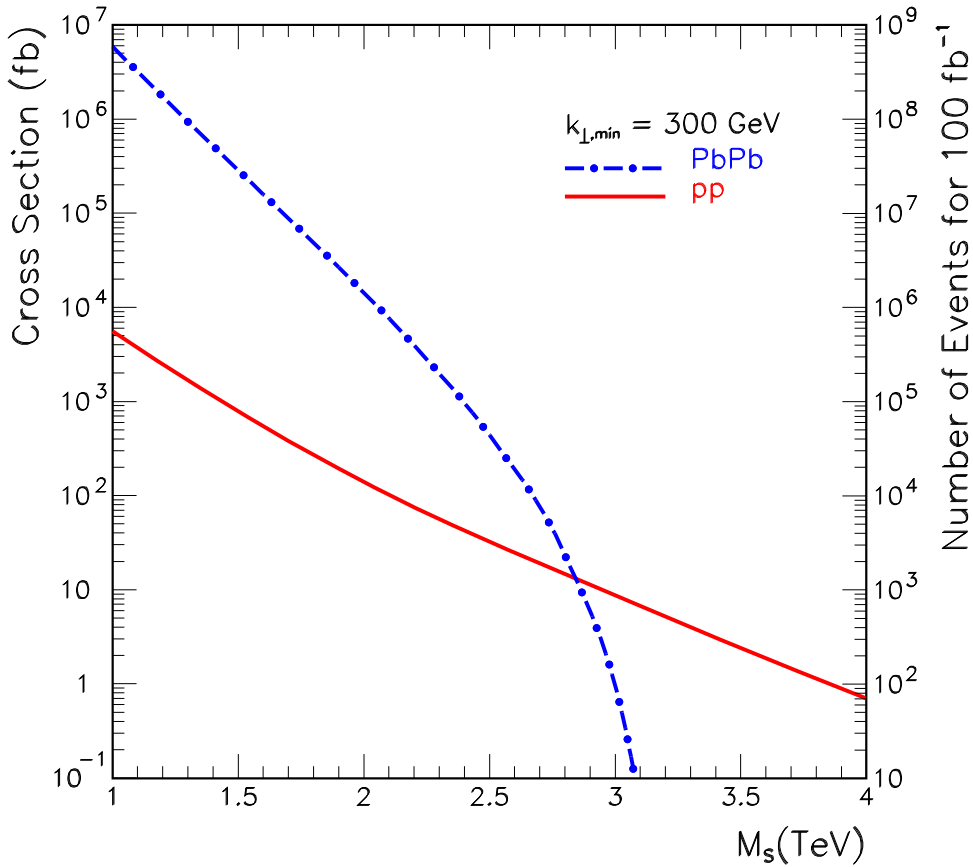}{0.8}
\caption{Cross section for gluon fusion into
$\gamma + {\rm jet}|_{k_{\perp}(\gamma)> 300~{\rm GeV}}$ and expected number of
  events, for $100~{\rm fb}^{-1}$ and varying string scale.}
\label{fig:sigma}
\end{figure}

\begin{figure}
 \postscript{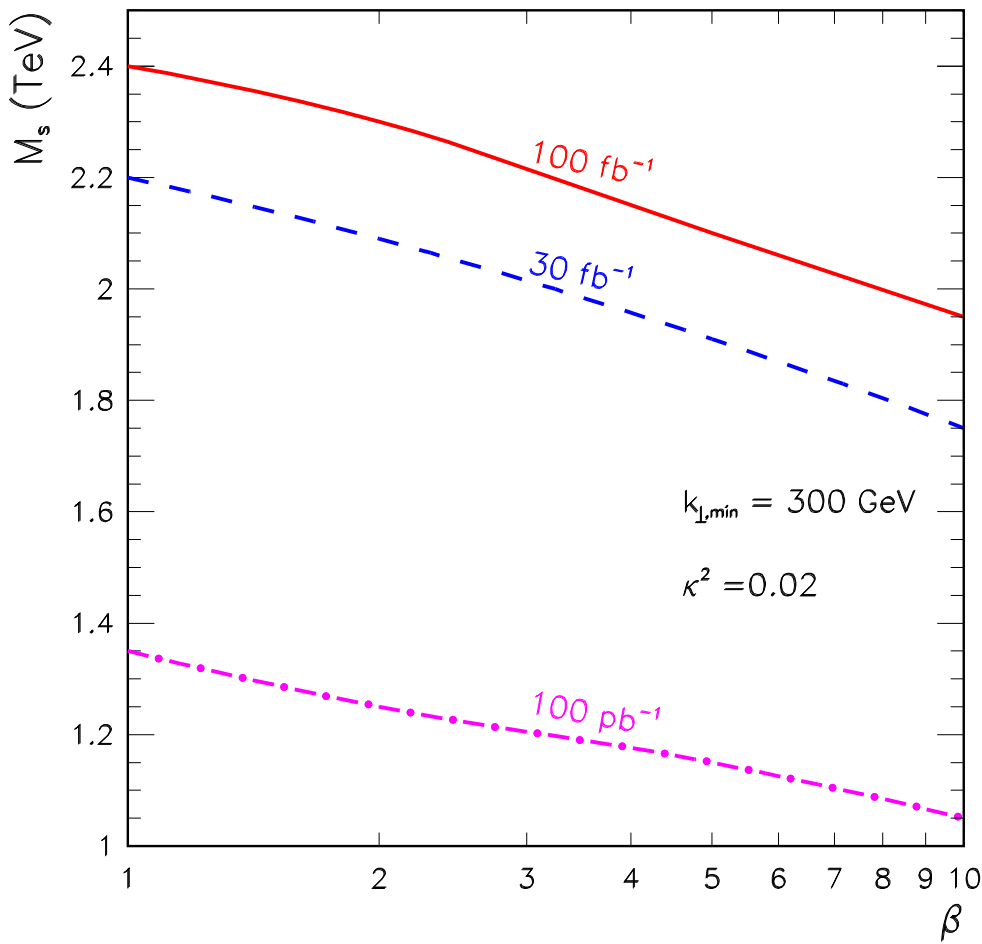}{0.8}
 \caption{Contours of 5$\sigma$ discovery in the ($M_s$, detector
   efficiency) plane for different integrated luminosities and
   $\kappa^2 =0.02$.}
\label{fig:S2N}
\end{figure}

In order to assess the possibility of discovery of signal above QCD
background, we adopt the kind of signal introduced
in~\cite{Dimopoulos:2001hw} to study detection of TeV-scale black
holes at the LHC, namely a high-$k_\perp$ isolated $\gamma$ or $Z.$
Thus, armed with parton distribution functions
(CTEQ6D)~\cite{Pumplin:2002vw} we have calculated integrated cross
sections $\sigma(pp\rightarrow \gamma + {\rm jet})|_{k_\perp (\gamma)>
  k_{\perp, {\rm min}}}$ for both the background QCD processes (see
Appendix I) and for $gg\rightarrow \gamma g$, for an array of values
for the string scale $M_s$ (see Appendix II). Our results are shown in
Fig.~\ref{fig:ptfig}. It is evident that the background is
significantly reduced for large $k_{\perp, {\rm min}}$. At very large
values of $k_{\perp, {\rm min}},$ however, event rates become
problematic. In Fig.~\ref{fig:sigma} we show the string cross section
and number of events (before cuts) in a 100 fb$^{-1}$ run at LHC, for
$k_{\perp, {\rm min}}=300$~GeV, as a function of the string scale
$M_s$. Next, we explore the LHC discovery potential by computing the
signal-to-noise ratio (${\rm signal}/\sqrt{\rm SM\ background} \equiv
{\rm S}/{\rm N}$). For a 300~GeV cut in the transverse momentum, the
QCD cross section (shown in Fig.~\ref{fig:ptfig}) is about $8 \times
10^3$~fb, yielding (for 100~fb$^{-1}$) $\sqrt{\rm SM\ background}
\approx 895.$ A point worth noting at this juncture: to minimize
misidentification with a high-$k_\perp \ \pi^0$, isolation cuts must
be imposed on the photon, and to trigger on the desired channel, the
hadronic jet must be identified~\cite{Bandurin:2003kb}.  We will leave
the exact nature of these cuts for the experimental groups, and
present results for a generous range of direct photon reconstruction
efficiency. To do so, we define the parameter
\begin{equation}
 \beta = \frac{{\rm background \, due \, to \, misidentified} \, \pi^0 \, {\rm
after \, isolation \, cuts}}{{\rm QCD \, background \, from \, direct \, photon \, production}}  + 1 \,\, .
\end{equation}
Therefore, the noise is increased by a factor of $\sqrt{\beta}$, over
the direct photon QCD contribution.  Our significant results are
encapsuled in Fig.~\ref{fig:S2N}, where we show the discovery reaches
of the LHC for different integrated luminosities and
$\kappa^2 = 0.02$.  A detailed study of the CMS potential for
isolation of prompt-$\gamma$'s has been recently carried
out~\cite{Gupta:2007cy}, using GEANT4 simulations of $\gamma + {\rm
  jet}$ events generated with Pythia. This analysis (which also
includes $\gamma$'s produced in the decays of $\eta$, $K_s^0$,
$\omega^0,$ and bremsstrahlung photons emerging from high-$p_\perp$
jets) suggests $\beta \simeq 2$. Of course, considerations
of detector efficiency further reduce the $S/N$ ratio by an additional
factor $\epsilon$, where $1 < \epsilon \ll \sqrt{\beta}$. We conclude
that {\em discovery at the LHC would be possible for $M_s$ as large as
  2.3~TeV.} The dependence of the discovery reach with the $C^0-Y$
mixing coefficient $\kappa$ has been extensively discussed in the
accompanying Letter~\cite{Anchordoqui:2007da}.

We now briefly explore the potential of ALICE to search for low mass
string excitations~\cite{Conesa:2007nx}. With this motivation, we
extend our analysis to include heavy ions collisions. In the spirit of
Ref.~\cite{Chamblin:2002ad} we consider the unshadowed parton
distribution functions, i.e.,
\begin{equation}
R_{i/A} (x) = \frac{f_{i/A} (x,Q)}{A f_{i} (x,Q)}
\simeq 1 \, ,
\end{equation}
where $f_{i/A}$ and $f_{i}$ are the parton distribution functions
inside a free nucleus of mass $A$ and free nucleon, respectively.  For
$M_s \agt 1$~TeV, this approximation holds because LHC Pb-Pb collisions
probe the minimum value of parton momentum at $x_{\rm min} \approx
M_s^2/s \sim 0.033,$ where there are no shadowing effects. A
comparison of the string cross section for gluon fusion into $\gamma +
{\rm jet}|_{k_{\perp}(\gamma)> 300~{\rm GeV}}$ for $pp$ and Pb-Pb
collisions is shown in Fig.~\ref{fig:sigma}. However, the larger aggregate of
partons also increase the SM background; namely, for $k_{\perp, {\rm
    min}} > 300~{\rm GeV},$ $\sigma_{{\rm Pb-Pb} \to \gamma X} \approx
2.8 \times 10^7~{\rm fb}.$ This greatly decreases the sensitivity to D-brane 
models, which would 
require a Pb-Pb integrated luminosity of a few hundred pb$^{-1}$. This 
is substantially larger than the present day estimate~\cite{Dainese:2007dg}.

\section{Bump-Hunting}
\label{s4}

\begin{figure}
 \postscript{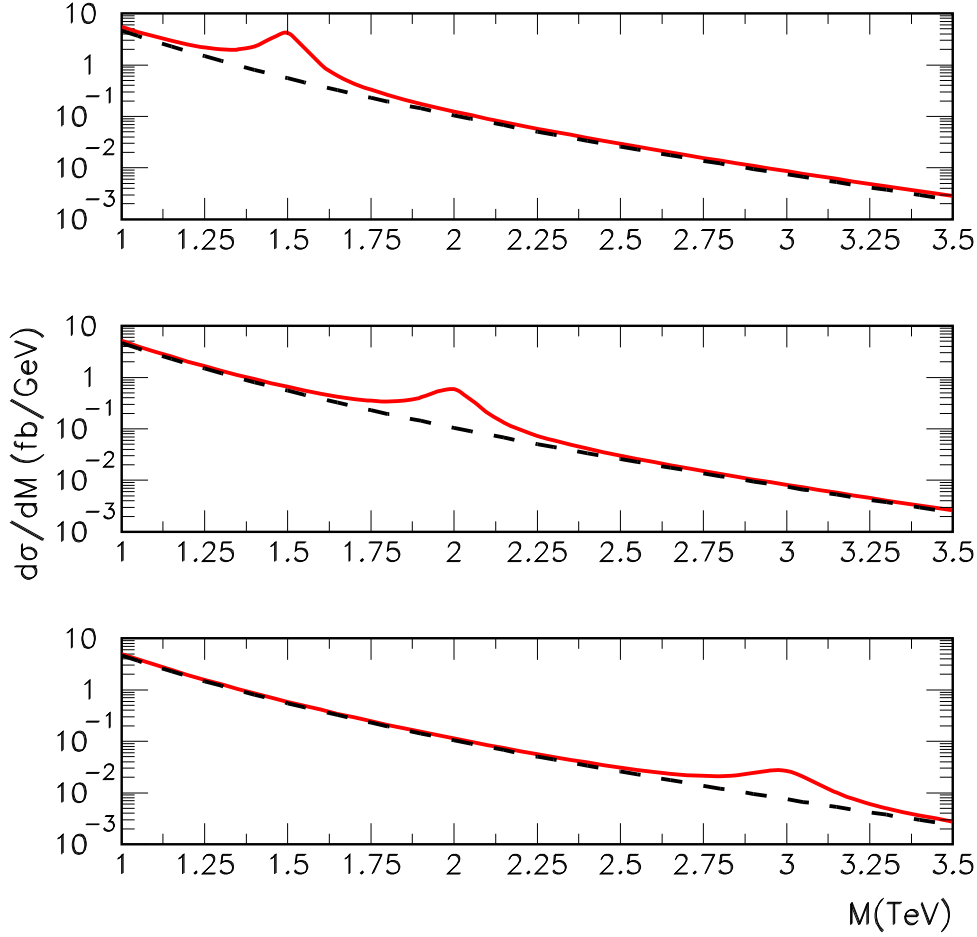}{0.8}
 \caption{$d\sigma/dM$ (units of fb/GeV) {\em vs.} $M$ (TeV) is plotted for 
the case of SM QCD background (dashed) and (first resonance)  string signal 
+ background (solid).}
\label{fig:bump}
\end{figure}

The discovery trigger described in the previous section, the
observation of isolated photons at large transverse momentum, serves
very well as a signature of new physics.
%However, as mentioned above,
%this criterion served also as a marker for
%Hawking radiation following
%production of TeV-scale black holes at
%LHC~\cite{Dimopoulos:2001hw}.
Given the particular nature of the
process we are considering, the production of a TeV-scale resonance
and its subsequent 2-body decay, signatures in addition to large
$k_\perp$ photons are available. Most apparently, one would hope that
the resonance would be visible in data binned according to the
invariant mass $M$ of the photon + jet, setting cuts on photon and jet
rapidities, $y_1,\, y_2 < y_{\rm max}$, respectively.  With the
definitions $Y\equiv \thalf (y_1 + y_2)$ and $y \equiv
\thalf(y_1-y_2)$, the cross section per interval of $M$ for
$pp\rightarrow \gamma + {\rm jet} +X$ is given
by~\cite{Eichten:1984eu}
\begin{eqnarray} \frac{d\sigma}{dM} & = & M\tau\ \sum_{ijk}\left[
\int_{-Y_{\rm max}}^{0} dY \ f_i (x_a,\, M)  \right. \ f_j (x_b, \,M ) \
\int_{-(y_{\rm max} + Y)}^{y_{\rm max} + Y} dy
\left. \frac{d\sigma}{d\hat t}\right|_{ij\rightarrow \gamma k}\ \frac{1}{\cosh^2
y} \nonumber \\
& + &\int_{0}^{Y_{\rm max}} dY \ f_i (x_a, \, M) \
f_j (x_b, M) \ \int_{-(y_{\rm max} - Y)}^{y_{\rm max} - Y} dy
\left. \left. \frac{d\sigma}{d\hat t}\right|_{ij\rightarrow \gamma k}\
\frac{1}{\cosh^2 y} \right]
\label{longBH}
\end{eqnarray}
where $i,j,k$ are different partons, $\tau = M^2/s$, $x_a =
\sqrt{\tau} e^{Y}$, and $x_b = \sqrt{\tau} e^{-Y}.$ The kinematics of
the scattering provides the relation
\begin{equation}
k_\perp = \frac{M}{2\, \cosh y}
\label{relation}
\end{equation}
which, when combined with the standard cut $k_\perp \agt k_{\perp,
  {\rm min}}$, imposes a {\em lower} bound on $y$ to be implemented in
the limits of integration. (For details see Appendix~III.)  The $Y$
integration range in Eq.~(\ref{longBH}), $Y_{\rm max} = {\rm min} \{
\ln(1/\sqrt{\tau}),\ \ y_{\rm max}\}$, comes from requiring $x_a, \,
x_b < 1$ together with the rapidity cuts $|y_1|, \, |y_2| \le
2.4$. Finally, the Mandelstam invariants occurring in the cross
section are given by $\hat s = M^2,$ $\hat t = -\thalf M^2\ e^{-y}/
\cosh y,$ and $\hat u = -\thalf M^2\ e^{+y}/ \cosh y.$

In Fig.~\ref{fig:bump} we show several representative plots of this
cross section for different values of $M_s$. Standard bump-hunting
methods, such as calculating cumulative cross sections
\begin{equation}
\sigma (M_0) = \int_{M_0}^\infty  \frac{d\sigma}{dM} \, \, dM
\end{equation}
and searching for regions with significant deviations from the QCD
background, may allow to find an interval of $M$ suspected of
containing a bump.  With the establishment of such a region, one may
calculate a signal-to-noise ratio, with the signal rate estimated in
the invariant mass window $[M_s - 2 \Gamma, \, M_s + 2 \Gamma]$. This
estimate of signal-to-noise would be roughly the same as that obtained
through the inclusive cut $k_\perp > 300$~GeV. This follows from the
relation~(\ref{relation}): for $M$ in the range of $M_s \agt 2$ and
for the significant contributing regions of $y$, the resulting
$k_\perp$ cut in Eq.~(\ref{relation}) does not differ significantly
from the 300~GeV~\cite{2pole}.  Should bumps be found, the D-brane model can be further
differentiated from other TeV-scale resonant processes by the details
of the angular distributions inherent in Eq.~(\ref{mhvlow2}).

\section{Concluding Discussion}
\label{s5}

In this work we have described how to search for the effects of Regge
excitations of fundamental strings at LHC collisions. The underlying
parton process for the excitation of the string resonance is
dominantly the single photon production in gluon fusion, $gg \to
\gamma g$, with open string states propagating in intermediate
channels. If the photon mixes with the gauge boson of the baryon
number, which is a common feature of D-brane quivers, the amplitude
appears already at the string disk level. It is completely determined
by the mixing parameter -- and it is otherwise
model-(compactification-) independent. We have shown that even for
relatively small mixing, 100~fb$^{-1}$ of LHC data (in the $pp \to
\gamma +$ jet channel) could probe deviations from SM physics at a
$5\sigma$ significance, for $M_s$ as large as 2.3~TeV. We note that
such a numerical value for the discovery reach
is lower than the estimate presented in  the accompanying Letter, 
$M_s \sim 3.3~{\rm TeV}$~\cite{Anchordoqui:2007da}. 
The present analysis contains a refined treatment of the resonance region,
including the recently computed decay widths of both spin $J=0$ and spin $J=2$ Regge recurrence of the gluon octet \cite{widths}. The discovery reach is lower because these resonances are slightly wider than na\"ively expected while
the signal cross sections are very
sensitive to the width values.

In closing we discuss some interesting contrast of $\gamma$ and $Z$
production that can serve as an additional marker of the D-brane
model.  Ignoring the $Z$-mass (i.e., keeping only transverse $Z$'s),
and assuming that cross sections $\times$ branching into lepton pairs
are large enough for complete reconstruction to $pp \to Z + {\rm
  jet},$ the quiver contribution to the signal is suppressed relative
to the photon signal by a factor of $\tan^2\theta_W = 0.29.$ The SM
ratio ($Z$ background)/( $\gamma$ background) is roughly 0.92 for
processes involving $u$ (or $\bar u$) quarks, and 4.7 for processes
involving $d$ (or $\bar d$) quark.  Thus, even if $d$ quark processes
are ignored, one obtains a signal-to-noise ratio $ ({\rm S}/{\rm
  N})_Z=0.29/\sqrt{0.92}=0.30 \, ({\rm S}/{\rm N}_\gamma).$ Keeping
the $d$ quarks will only lead to more suppression of $({\rm S}/{\rm
  N})_Z$~\cite{systematics}. This implies that if the high-$k_\perp$
photons, as predicted by the TeV string model, are discovered at
$5\sigma,$ they will not be accompanied by any significant deviation
of $pp \to Z + {\rm jet}$ from SM predictions. This differs radically
from the evaporation of black holes produced at the LHC.  In such a
case, production of high-$k_\perp$ $Z$ and $\gamma$ are comparable.
The suppression of high-$k_\perp$ $Z$ production, whose origin lies in
the particular structure of the quiver model, will hold true for all
the low-lying levels of the string.

\section*{Acknowledgments}

L.A.A.\ is supported by the U.S. National Science Foundation and the
UWM Research Growth Initiative.  H.G.\ is supported by the
U.S. National Science Foundation Grants No PHY-0244507 and PHY-0757959.  The research
of T.R.T.\ is supported by the U.S.  National Science Foundation Grants
PHY-0600304, PHY-0757959 and by the Cluster of Excellence ``Origin and Structure of
the Universe'' in Munich, Germany.  He is grateful to Dieter L\"ust
and to Max--Planck--Institut f\"ur Physik,
Werner--Heisenberg--Institut in Munich for their kind hospitality.
Any opinions, findings, and
conclusions or recommendations expressed in this material are those of
the authors and do not necessarily reflect the views of the National
Science Foundation.

\section*{Appendix I}

The SM background for processes with a single photon in the final
state originates in the parton tree level processes $g q \rightarrow
\gamma q,\ g\bar q\rightarrow \gamma\bar q\ {\rm and} \ q\bar
q\rightarrow \gamma g$,
\begin{eqnarray}
\left. 2 E' \frac{d\sigma}{d^3k'} \right|_{pp \to \gamma X} & = &  \sum_{ijk}
\left. \int dx_a \, dx_b \, f_i(x_a,Q) \, f_j (x_b,Q) \, 2 E' \frac{d \hat \sigma}{d^3k'}\right|_{ij \to \gamma k} \ ,
\end{eqnarray}
where $x_a$ and $x_b$ are the fraction of momenta of the parent
hadrons carried by the partons which collide, $k'$ $(E')$ is the
photon momentum (energy), $d \hat \sigma/d^3k'|_{ij \to \gamma k}$ is
the cross section for scattering of partons of type $i$ and $j$
according to elementary QCD diagrams, $f_i(x_a,Q)$ and $f_j (x_b, Q)$
are parton distribution functions, $Q$ is the momentum transfer, and
the sum is over the parton species: $g, q = u,\ d,\ s,\
c,\ b$. In what follows we focus on $gq \to \gamma q$, which results
in the dominant contribution to the total cross section. Corrections
from the other two processes can be computed in a similar fashion.
The hard parton-level cross section reads,
\begin{equation}
  \left. 2 E' \frac{d \hat \sigma}{d^3k'} \right|_{gq \to \gamma q}  =  \frac{(2 \pi)^4}{(2 \pi)^6} \, \frac{1}{2 \hat s} \, \delta[(k+p-k')^2]
  \, \frac{1}{4}\sum |{\cal M}|^2
  =  \frac{1}{(2\pi)^2} \frac{1}{2 \hat s} \, \delta(2p \ . \ q + q^2) \frac{1}{4}\sum |{\cal M}|^2 \,,
\label{under}
\end{equation}
where $k$ and $p$ are the momenta of the incoming partons, $q = k - k',$ $\hat s = x_a\, x_b\, s$, and  $-q^2 = -\hat t = Q^2.$ Here,
\begin{equation}
 \frac{1}{4} \sum |{\cal M}|^2\, = \frac{1}{3} g^2 e^2 e_q^2 \left(\frac{\hat s}{\hat s + \hat t} + \frac{\hat s + \hat t}{\hat s} \right),
\end{equation}
where $g$ and $e$ are the QCD and electromagnetic coupling constants,
and $e_q$ is the fractional electric charge of species $q$. For
completeness we note that for $q \bar q \to g \gamma$,
\begin{equation}
 \frac{1}{4} \sum |{\cal M}|^2 \, = \frac{8}{9} g^2 e^2 e_q^2 \left(-
\frac{\hat t}{\hat s + \hat t} - \frac{\hat s + \hat t}{\hat t} \right)\, .
\end{equation}
Equation~(\ref{under}) can be
most conveniently integrated in terms of the rapidity $y$ and
transverse momentum $k_\perp$ of the final photon
\begin{equation}
\frac{d^3k'}{2E'} = \frac{1}{2}  d^2k_\perp \, dy = \pi k_\perp\,
dk_\perp\, dy\, .
\end{equation}
Considering that the incoming momentum of the gluon is $k = x_a P_1$
and that of the quark is $p = x_b P_2$, we can re-write
the argument of the delta function as
\begin{equation}
2 p \ . \ q + q^2 = 2\, x_b \, P_2 \ . \ (x_a P_1 - k') + \hat t = x_a\, x_b\, s - 2\, x_b \, P_2 \ . \ k' + \hat t \, \, ,
\end{equation}
where $P_1$ and $P_2$ are the initial momenta of the parent protons.
Introducing, $k'_0 = k_\perp \, \cosh y$, $k'_\parallel = k_\perp \, \sinh y$, $P_1 = (\sqrt{s}/2,\, 0,\, 0,\, \sqrt{s}/2)$, and $P_2 = (\sqrt{s}/2,\, 0,\, 0,\, -\sqrt{s}/2)$ we obtain
\begin{equation}
P_2 \ . \ k' = \frac{\sqrt{s}}{2} \, k_\perp (\cosh y + \sinh y) =
\frac{\sqrt{s}}{2} \, k_\perp \, e^y
\end{equation}
and
\begin{equation}
\hat t = - 2 k \ .\ k' = - 2 x_a \frac{\sqrt{s}}{2} \, k_\perp \, e^{-y} = - \sqrt{s} \, k_\perp\, e^{-y} \, x_a \,,
\end{equation}
so that
\begin{eqnarray}
\delta (x_a\, x_b\, s - \sqrt{s}\, x_b\, k_\perp \, e^y - \sqrt{s} \, x_a \, k_\perp \, e^{-y})
& = & \frac{1}{s} \, \, \delta ( x_a \, x_b - x_b \,x_\perp \, e^y - x_a \, x_\perp \, e^{-y}) \nonumber \\
 & = & \frac{1}{s \, \left[x_a - x_\perp \, e^{y}\right]} \, \, \, \delta \left(x_b - \frac{x_a \, x_\perp \, e^{-y}}{x_a - x_\perp \, e^y}\right) \,,
\end{eqnarray}
where $x_\perp = k_\perp /\sqrt{s}.$ The lower bound $x_b > 0$ implies $x_a > x_\perp \, e^y$. The upper bound $x_b < 1$ leads to a stronger constraint
\begin{equation}
x_a > \frac{x_\perp e^y}{1 - x_\perp e^{-y}} \,,
\label{bound}
\end{equation}
which requires $x_\perp e^y < 1 - x_\perp e^{-y}$, yielding $x_\perp <
(2 \, {\rm cosh}\, y)^{-1}$. Of course there is another completely
symmetric term, in which $g$ comes from $P_2$ and $q$ comes from
$P_1$.  Putting all this together, the total contribution from $gq \to
\gamma q$ reads
\begin{eqnarray}
  \sigma_{pp \to \gamma X}^{qg \to \gamma q} & = &  2\, \sum_q \int \frac{d^3k'}{2E'} \int dx_a \int dx_b \, f_g(x_a,Q) \, f_{q}(x_b,Q) \, \frac{1}{(2\pi)^2} \, \,
 \frac{1}{s \, \left[x_a - x_\perp e^y \right]} \nonumber \\
 & \times &  \frac{1}{2 \hat s} \, \, \delta \left(x_b - \frac{x_a x_\perp e^{-y}}{x_a - x_\perp e^{y}} \right) \, \, \frac{e^2 g^2 e_q^2}{3} \, \, \left(\frac{\hat s + \hat t}{\hat s} + \frac{\hat s}{\hat s + \hat t} \right) \, .
\label{laven}
\end{eqnarray}
With the change of variables $z = e^y$ Eq.~(\ref{laven}) can be re-written as
\begin{eqnarray}
 \sigma_{pp \to \gamma X}^{qg \to \gamma q} & = & 2\, \sum_q \int \frac{\pi \, k_\perp \, dk_\perp \, dz}{z} \int dx_a \int dx_b \, f_g(x_a,Q) \, f_{q} (x_b,Q) \frac{1}{(2 \pi)^2 \, 2 x_a \, x_b \, s^2 (x_a - x_\perp z)} \nonumber \\
& \times & \delta \left(x_b - \frac{x_a x_\perp z^{-1}}{x_a - x_\perp z}\right)  \frac{e^2 g^2 e_q^2}{3} \left(\frac{\hat s + \hat t}{\hat s} + \frac{\hat s}{\hat s + \hat t} \right) \,\, .
\label{esta}
\end{eqnarray}
Now, since
\begin{equation}
\frac{\hat t}{\hat s} = - \frac{\sqrt{s} k_\perp e^{-y}}{x_b s} = -\frac{x_\perp}{x_b \, z} = \frac{x_\perp \, z}{x_a} - 1 \, ,
\end{equation}
Eq.~(\ref{esta}) becomes
\begin{eqnarray}
 \sigma_{pp \to \gamma X}^{qg \to \gamma q} & = & \frac{e^2 g^2}{12 \pi s}\, \int_{x_{\perp {\rm min}}}^{1/2} dx_\perp \, \int_{z_{\rm min}}^{z_{\rm max}}  dz \int_{x_{a,min}}^1 dx_a\,\, f_g(x_a,Q) \left[\sum_q e_q^2\,\, f_{q}\left( \frac{ x_a x_\perp z^{-1}}{x_a - x_\perp z},Q \right)
\right] \nonumber \\
 & \times & \frac{1}{x_a^2} \, \left(\frac{x_\perp z}{x_a} + \frac{x_a}{x_\perp z}\right) \,,
\label{hc}
\end{eqnarray}
where the integration limits,
\begin{equation}
z_{^{\max}_{\rm min}} = \frac{1}{2} \left[ \frac{1}{x_\perp} \pm \sqrt{\frac{1}{x_\perp^2} -4} \right] \ \ \ \ \ \ \ \ \ \ \ \ {\rm and} \ \ \ \ \ \ \ \ \ \ \ \
x_{a,{\rm min}} = \frac{x_\perp z}{1 - x_\perp z^{-1}} \, ,
\end{equation}
are obtained from Eq.~(\ref{bound}). In Fig.~\ref{fig:ptfig} we show
the QCD background cross section {\em vs} $k_{\perp, {\rm min}}$, as
obtained through numerical integration of Eq.~(\ref{hc}).  To
accommodate the minimal acceptance cuts on final state photons from the
CMS and ATLAS proposals~\cite{Ball:2007zza}, an additional kinematic
cut, $|y|<2.4,$ has been included in the calculation.

\section*{Appendix II}

For the considerations in the present work, the resonant cross section
can be safely approximated by single poles in the Narrow-Width
Approximation,
\begin{equation}
\frac{\Gamma \sqrt{s_0}/\pi}{(\hat s - s_0)^2 + (\Gamma \sqrt{s_0})^2}\, \frac{\pi}{\Gamma \sqrt{s_0}} =  \frac{\pi}{\Gamma \sqrt{s_0}} \,\, \delta(\hat s - s_0) \,,
\end{equation}
where $s_0 = M_s^2$.  The scattering proceeds through $J=0$ and $J=2$
angular momentum states, with the $M_s^8$ term in Eq.~(\ref{mhvlow2})
originating from $J=0$, and the $t^4+ u^4$ piece reflecting $J=2$
activity. The widths of these two resonances are different, with
$\Gamma^{J=0} = (3/4) \, \alpha_s M_s,$ and $\Gamma^{J=2}= (9/20)\,
\alpha_s M_s$~\cite{widths}.  The average string amplitude square in
Eq.~(\ref{mhvlow2}) then becomes
\begin{eqnarray}
|{\cal M}(gg\to g\gamma)|^2 & \approx &
4g^4Q^2C(N) \, \frac{\pi}{s_0^{5/2}}
\, \left[\frac{s_0^4}{\Gamma^{J=0}}+ \frac{\hat t^4+ (\hat t + s_0)^4}{
\Gamma^{J=2}} 
\right] \, \delta(\hat s - s_0) \nonumber \\
 & = & 4g^4Q^2C(N) \, \frac{\pi}{\alpha_s \, s_0^{3}}
\, \left\{\tfrac{4}{3} s_0^4+ \tfrac{20}{9} [\hat t^4+ (\hat t + s_0)^4] 
\right\} \, \delta(\hat s - s_0) \, . 
\end{eqnarray}
Thus, the total cross section for single photon
production in gluon fusion is given by
\begin{eqnarray}
  \sigma_{pp \to \gamma X}^{gg \to \gamma g} & = &  \int \frac{d^3k'}{2E'} \int dx_a \int dx_b \, f_g(x_a,Q) \, f_{g}(x_b,Q) \, \frac{1}{(2\pi)^2} \, \,
 \frac{1}{2\,\hat s\, s} \delta(x_a\, x_b -x_b x_\perp z - x_a x_\perp z^{-1}) \nonumber \\
 & \times & 4 g^4 Q^2 C(N)  \frac{\pi}{\alpha_s \, s_0^3} \, \left\{ 
\tfrac{4}{3} s_0^4 + \tfrac{20}{9} [\hat t^4 + (\hat t + s_0)^4 ] \right\}\, \delta (\hat s - s_0) \,.
\label{mel}
\end{eqnarray}
We set $Q = M_s$, which is appropriate for the dual picture of string
theory. We are aware that for $Q \sim M_s$, the parton distribution
functions will receive significant corrections from the rapid increase
of degrees of freedom. Fortunately, as noted
elsewhere~\cite{Anchordoqui:2001cg}, at parton center-of-mass energies
corresponding to low-lying string excitations the resonant cross
section is largely insensitive to the details of the choice of $Q$.
Plugging $\tau_0 = s_0/s$ into Eq.~(\ref{mel}), we obtain
\begin{eqnarray}
  \sigma_{pp \to \gamma X}^{gg \to \gamma g} & = &  \int \frac{\pi \, k_\perp \, dk_\perp \, dz}{z} \int dx_a \int dx_b \, f_g(x_a,Q) \, f_{g}(x_b,Q) \,
 \, \frac{\delta(x_a\, x_b -x_b x_\perp z - x_a x_\perp z^{-1})}{
8\, \pi^2 \,  x_a^2\, x_b} \nonumber \\
 & \times & 4 g^4 Q^2 C(N) \frac{\pi \, s}{\alpha_s \, s_0^3} \, \left\{ \tfrac{4}{3} \tau_0^4 +  \tfrac{20}{9} [(x_a \, x_\perp z^{-1})^4 + (-x_a\, x_\perp \, z^{-1} + \tau_0)^4] \right\} \nonumber \\
 & \times & \delta\left(x_b - \frac{\tau_0}{x_a}\right) \,,
\end{eqnarray}
which after integration over $x_b$ leads to
\begin{eqnarray}
  \sigma_{pp \to \gamma X}^{gg \to \gamma g} & = & \frac{g^4 Q^2 C(N)}{2\, \alpha_s \tau_0^4 s} \int \frac{x_\perp \, dx_\perp \, dz}{z} \int dx_a \, f_g(x_a,Q) \, f_{g}(\tau_0/x_a,Q) \,\frac{1}{x_a} \nonumber \\
 & \times &\,\,
 \delta \left(\tau_0 - \frac{\tau_0 x_\perp z}{x_a} - \frac{x_a x_\perp}{ z}\right)   \, \left\{ \tfrac{4}{3} \tau_0^4 + \tfrac{20}{9} [ (x_a \, x_\perp z^{-1})^4 + (-x_a\, x_\perp \, z^{-1} + \tau_0)^4] \right\} .
\end{eqnarray}
\begin{figure}
 \postscript{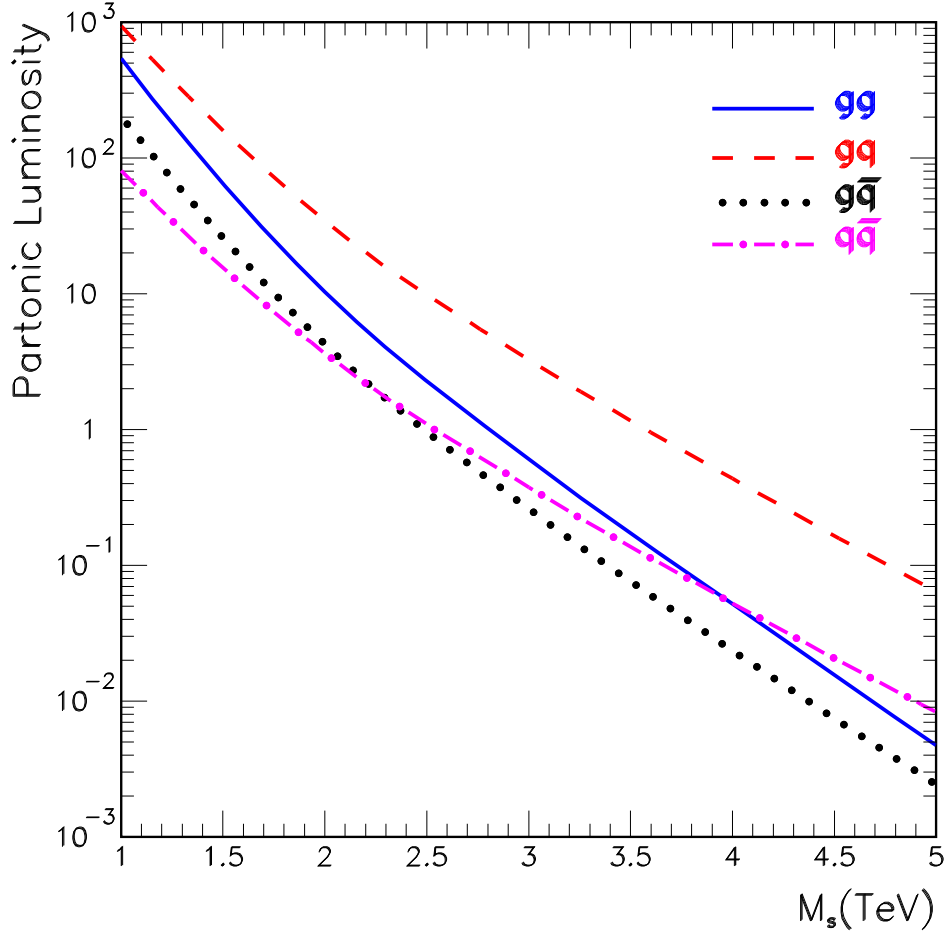}{0.78}
 \caption{Relative contributions of initial state partons ($ij = gg,\ gq,\ g\bar q,\ {\rm and}\ q \bar q$) to $\int_{\tau_0}^1 f_i(x_a, Q) \, \, f_j(\tau_0/x_a, Q)\,\,\, dx_a/x_a$, with varying string scale.}
\label{fig:pl}
\end{figure}
We now make use of the delta function scaling property,
\begin{equation}
\delta \left(\tau_0 - \frac{\tau_0 x_\perp z}{x_a} - \frac{x_\perp x_a}{z} \right) = \delta(f(z)) = \frac{1}{|f'(z_+)|} \, \, \delta (z - z_+) + \frac{1}{|f'(z_-)|} \, \, \delta (z - z_-) \,,
\end{equation}
where $z_\pm$ are the solutions to $f(z) = 0$,
\begin{equation}
z_\pm = \frac{x_a}{2 x_\perp}\, \left( 1 \pm \sqrt{1 - \frac{4 x_\perp^2}{\tau_0}} \right) \, \,.
\end{equation}
Besides,
\begin{equation}
\frac{1}{z_\pm |f'(z_\pm)| }= \left| \frac{\tau_0 x_\perp z_\pm}{x_a} - \frac{x_a \, x_\perp}{z_\pm} \right|^{-1}
\end{equation} 
and
\begin{equation}
\frac{x_a \, x_\perp}{ z_\pm}  = \frac{\tau_0}{2} \,\left (1 \mp
\sqrt{1 - \frac{4 x_\perp^2}{\tau_0}} \right) \,;
\end{equation} 
therefore,
\begin{equation}
\frac{1}{z_\pm |f'(z_\pm)|} = \frac{1}{\tau_0\,
\sqrt{1 - 4 x_\perp^2/ \tau_0}} \, .
\end{equation}
A straightforward calculation shows that
\begin{eqnarray}
\tfrac{16}{9} \, \tau_0^2 \, (5 \, x_\perp^4  -10 \, x_\perp^2\, \tau_0  
+ 4 \, \tau_0^2) 
& = &
\left\{\tfrac{4}{3}\, \tau_0^4 + \tfrac{20}{9} \, [(x_a x_\perp z_+^{-1})^4 + (-x_a x_\perp z_+^{-1} + \tau_0)^4] \right\} \nonumber \\ & + &
\left\{ \tfrac{4}{3} \tau_0^4 + \tfrac{20}{9} \, [ (x_a x_\perp z_-^{-1})^4 + (-x_a x_\perp z_-^{-1} + \tau_0)^4] \right\} \,,
\end{eqnarray}
and hence integration over the $z$ variable yields
\begin{eqnarray}
  \sigma_{pp \to \gamma X}^{gg \to \gamma g} & = & \frac{8}{9} \, \frac{g^4 \, Q^2 C(N)}{\alpha_s \, \tau_0^3 \, s} \, \int_{x_{\perp, {\rm min}}}^{\sqrt{\tau_0}/2} d x_\perp \, \frac{x_\perp}{\sqrt{1- 4x_\perp^2/\tau_0}} \,\, \left(5 \,  x_\perp^4 - 10 \, x_\perp^2 \, \tau_0 + 4 \, \tau_0^2 \right)  \nonumber \\
 & \times & \int_{\tau_0}^1 \frac{dx_a}{x_a} \, \, f_g(x_a,Q) \, \, f_g (\tau_0/x_a,Q)  \, ,
\end{eqnarray}
where the integration range has been derived from the conditions
$0 < x_b = \tau_0/x_a < 1$ and $4 x_\perp^2<\tau_0$, which imply $\tau_0 < x_a < 1$ and
$x_{\perp, {\rm min}} < x_\perp < \sqrt{\tau_0}/2$.
Finally, integration over $x_\perp$ leads to
\begin{eqnarray}
  \sigma_{pp \to \gamma X}^{gg \to \gamma g} & = & \frac{1}{9} \frac{g^4\, Q^2 C(N)}{\alpha_s \ \tau_0^2 \, s} \, \sqrt{1 - \frac{4 x_{\perp, {\rm min}}^2}{\tau_0}} \left(5\  \tau_0^2 - 6\  \tau_0 \ x_{\perp, {\rm min}}^2 + 2 \ x_{\perp, {\rm min}}^4\right) \nonumber \\
 & \times & \int_{\tau_0}^1 \frac{dx_a}{x_a} \, \,  f_g(x_a,Q) \, \, f_g (\tau_0/x_a,Q)  \, .
\label{csb}
\end{eqnarray}
Note that all stringy corrections to the pure bosonic cross section
given by Eq.~(\ref{csb}) have similar factorizations. An illustration of the
relative partonic luminosities of the different processes is shown in
Fig.~\ref{fig:pl}.

\section*{Appendix III}

We follow the same conventions and notation given in Appendix I for
two-body processes leading to final states consisting of $\gamma + \,
{\rm jet}$, with equal and opposite transverse momenta $k_\perp$ and
$p_\perp$, respectively. The distribution of invariant masses 
$M^2=(k' + p')^2$ is given by
\begin{eqnarray}
\frac{d\sigma}{dM^2}  & = &\frac{(2 \pi)^4}{(2 \pi)^6}\, 
\int \frac{d^3k'}{2E'_1}\, \int \frac{d^3p'}{2E'_2} \,  \sum_{ijk} 
\int dx_a \, \int dx_b\, f_i(x_a,M) \,f_j(x_b,M)\, \delta^4 (p - k' - p') 
\nonumber \\
 & \times & \delta(p^2 - M^2) \, \frac{1}{2 \hat s} \, 
\overline{\sum_{\rm spins} |{\cal M}|^2} \,\,,
\label{surf}
\end{eqnarray}
where
\begin{equation}
\overline{\sum_{\rm spins} |{\cal M}|^2} = |{\cal M}(ij \to \gamma k) |^2 = 64 \pi^2 \hat s \, \frac{d\sigma}{d\Omega} = 16 \pi \hat s^2 \, 
\left. \frac{d\sigma}{d\hat t} \right|_{ij \to \gamma k} \,,
\end{equation}
$p^2 = \hat s = (k'+p')^2 = 2 k'. \, p' = 2 E'_1 E'_2 - k'_\parallel p'_\parallel +
p_\perp^2,$ and 
\begin{equation}
\delta^4 (p - k'_\perp - p'_\perp) = \delta(E - E_1 -E_2)\, 
\delta(p_\parallel - k'_\parallel - p'_\parallel) \, 
\delta(\vec k_\perp + \vec p_\perp) \,\, .
\end{equation}
The integration over $d^3k'\,d^3p'$ can be conveniently re-written in terms 
of rapidities $y_1$ and 
$y_2$ (of the $\gamma$ and the ${\rm jet}$) and their common transverse 
momentum,
\begin{equation}
\frac{d^3p}{2E} = \frac{\pi}{2} \,\, dp^2_\perp \,\, dy \,,
\end{equation}
where $y \equiv \tfrac{1}{2} (y_1 - y_2)$.
Since $E'_1 = p_\perp \cosh y_1$, $k'_\parallel = p_\perp \sinh y_1,$
$E'_2 = p_\perp \cosh y_2$, and $p'_\parallel = p_\perp \sinh y_2,$ a 
straightforward calculation leads to $E'_1 E'_2 - k'_\parallel p'_\parallel = 
p_\perp^2 \, \cosh(y_1-y_2) \equiv p_\perp^2 \cosh 2y$.  Now, using the 
identity of hyperbolic functions, 
$1+ \cosh 2y  = 2 \cosh ^2y,$ we define
\begin{equation}
\tau = \frac{\hat s}{s} = \frac{M^2}{s} = \frac{4 p_\perp^2}{s} \cosh^2 y 
\end{equation}
so that
\begin{equation}
\delta(\hat s - M^2) = \delta(4 p_\perp^2 \cosh^2 y - M^2) = \frac{1}{4 \cosh^2 y}\ \  \delta\left(p_\perp^2 - \frac{M^2}{4 \cosh^2 y} \right) \,\, .
\end{equation}
Using 
\begin{equation}
\int d^2 \vec k_\perp \,\, d^2 \vec p_\perp \,\, 
\delta(\vec k_\perp + \vec p_\perp) \,\, \delta (p^2_\perp - M^2/4\cosh^2 y)
= \pi \int dp_\perp^2 \,  \delta (p^2_\perp - M^2/4\cosh^2 y) = \pi \,\,,
\end{equation}
Eq.~(\ref{surf}) becomes
\begin{eqnarray}
\frac{d\sigma}{dM^2} & = & \frac{\pi}{(2\pi)^2} \, \frac{1}{4} \, (8 \pi M^2) \int
dy_1 \, \int dy_2 \, \sum_{ijk} \int dx_a \, \int dx_b \  f_i(x_a,M) \ f_j(x_b,M) \ \frac{1}{4 \cosh^2 y}  \nonumber \\
 & \times &  \delta(E-E'_1-E'_2) \ \delta (p_\parallel - k'_\parallel - p'_\parallel) \  \left. 
\frac{d\sigma}{d\hat t} \right|_{ij \to \gamma k} \,\, .
\label{ibu}
\end{eqnarray}
We now define $a= E - E_1 - E_2$ and $b=p_\parallel - k'_\parallel -
p'_\parallel$ to perform the change of variables $A= a+b$ and $B =
a-b$, such that $\delta(a) \delta(b) = N \delta(A)\, \delta (B),$ with
normalization $N$ given by
\begin{equation}
\int da \, db \, \delta(a) \, \delta(b) = \int dA \, dB \, \frac{\partial(a,b)}{\partial (A,B)} \, N \, \delta(A) \, \delta(B) = \frac{N}{2} = 1 \, .
\end{equation}
The new variables can then be explicitly written as 
$\left\{^A_B \right\} = E \pm p_\parallel - (E_1 \pm k'_\parallel) - (E_2 \pm p'_\parallel)$, where $E\pm p_\parallel = \left\{^{\sqrt{s} x_a}_{\sqrt{s} x_b}\right\}$, $E_1 \pm k'_\parallel = p_\perp e^{\pm y_1} = p_\perp e^{\pm (Y +y)},$ and
$E_2 \pm p'_\parallel = p_\perp  e^{\pm y_2} = p_\perp e^{\pm (Y - y)}$, 
with $Y = \frac{1}{2} (y_1 + y_2)$.
Putting all this together, the product of delta functions in Eq.~(\ref{ibu}) 
becomes
\begin{eqnarray}
\delta(E-E_1-E_2) \ \delta(p_\parallel - k'_\parallel - p'_\parallel) & = & 2 \delta (\sqrt{s} x_a - 2 p_\perp e^Y \cosh y) \ \delta (\sqrt{s} x_b - 2 p_\perp e^{-Y} \cosh y)  \nonumber \\
 & = & 2 \delta(\sqrt{s} x_a - M e^Y)\, \delta(\sqrt{s} x_b - Me^{-Y}) \,,
\end{eqnarray}
and hence integration over the fraction of momenta is straightforward,
yielding
\begin{equation}
\frac{d\sigma}{dM} = \frac{1}{2} \, M \, \tau \, \int dy_1 \, dy_2 \ 
\frac{1}{\cosh^2 y} \ \sum_{ijk} f_i(\sqrt{\tau} e^Y,M) \,\, 
f_j(\sqrt{\tau} e^{-Y},M) \, \left. \frac{d\sigma}{d\hat
t}\right|_{ij\rightarrow \gamma k} \, \, .
\end{equation}
Now, if we constrain  the rapidities to the interval 
$2.4 < y_1, y_2 < 2.4$ we obtain the invariant mass spectrum given in Eq.~(\ref{longBH}). In addition, note that $x_a, x_b <1 $, implying $-\ln (1/\sqrt{\tau}) < Y < \ln (1/\sqrt{\tau})$. Besides, the cut on the transverse momentum leads to $k_{\perp, {\rm min}} < M /2 \cosh y.$ Finally, the Jacobian reads
\begin{equation}
dy_1\, dy_2 = \frac{\partial (y_1, y_2)}{\partial (Y,y)} dY dy = 2 \, dY \, dy 
\,,
\end{equation}
and the region of integration is defined by $|y_1| = |y+Y| < 2.4$ and $|y_2| 
= |y-Y| < 2.4$.


\begin{thebibliography}{99}
\bibitem{2pole} For $M_s < 1.5$~TeV, the second resonance of the
string mass spectrum ($\sqrt{n} M_s$, $n = 1,\ 2,\ \dots$) may be
observed at the LHC in the $pp \to \gamma + {\rm jet}$ channel as
well.
\bibitem{systematics} It is worth pausing to note that $\pi^0$
  misidentification does not play a role in the $Z$ channel, and so
  this tends to decrease the QCD background. On the other hand, the
  string signal will suffer some suppression because of finite mass
  effects. These systematics (which have opposite effects on $(S/N)_Z$)
  were not consider in the preceding discussion.
\end{thebibliography}
\end{document}